\documentclass[twocolumn,showpacs,preprintnumbers,aps,pre,nofootinbib]{revtex4-1}

\usepackage{amsmath}
\usepackage{amssymb}
\usepackage{graphicx}
\usepackage{epstopdf}
\usepackage{xspace}
\usepackage{hyperref}
\usepackage{color}

\DeclareMathOperator{\Prob}{Pr}

\newcommand{\SJT}{SJT\xspace}
\newcommand{\DW}{D-W Model\xspace}
\newcommand{\JA}{J-A Model\xspace}
\newcommand{\BA}{B-A network\xspace}

\begin{document}

\title{Social Judgment Theory Based Model On Opinion Formation, Polarization
 And Evolution}

\author{H.~F. Chau}\email{hfchau@hku.hk}
\author{C.~Y. Wong}
\author{F.~K. Chow}
\author{Chi-Hang Fred Fung}
\affiliation{Department of Physics and Center of Theoretical and Computational
 Physics, University of Hong Kong, Pokfulam Road, Hong Kong}

\date{\today}

\begin{abstract}
 The dynamical origin of opinion polarization in the real world is an
 interesting topic physical scientists may help to understand.
 To properly model the dynamics, the theory must be fully compatible with
 findings by social psychologists on microscopic opinion change.
 Here we introduce a generic model of opinion formation with homogeneous agents
 based on the well-known social judgment theory in social psychology by
 extending a similar model proposed by Jager and Amblard.
 The agents' opinions will eventually cluster around extreme and/or moderate
 opinions forming three phases in a two-dimensional parameter space that
 describes the microscopic opinion response of the agents.
 The dynamics of this model can be qualitatively understood by mean-field
 analysis.
 More importantly, first-order phase transition in opinion distribution is
 observed by evolving the system under a slow change in the system parameters,
 showing that punctuated equilibria in public opinion can occur even in a fully
 connected social network.
\end{abstract}



\maketitle

\section{Introduction}
\label{Sec:Intro}
 Opinion formation and evolution are interesting and important subject of
 research in social psychology.
 Many experiments and theories have been conducted and
 proposed~\cite{*[{See, for example, }][{, in particular chap.~8.}]Eagly:1993,%
 Griffin:2008}, including the elaboration likelihood model, the
 heuristic-systematic model and the cognitive dissonance theory.
 In particular, Sherif \emph{et al.} proposed the well-known social judgment
 theory (\SJT)~\cite{Eagly:1993,Sherif:1961,Sherif:1965,Griffin:2008}
 in the 1960's to explain the microscopic behavior of how individuals evaluate
 and change their opinions based on interaction with others.

 The basic idea of \SJT is that attitude change of an individual is a
 judgmental process.
 According to \SJT, describing the stand of an individual as a point in a
 continuum of possible opinions is not adequate because the individual's degree
 of tolerance is also important in determining his/her response to external
 stimuli and persuasion~\cite{Sherif:1961,Sherif:1965}.
 In particular, a presented opinion is acceptable (unacceptable) to a person if
 it is perceived to be sufficiently close to (far from) his/her own stand
 point.
 This presented opinion is said to be in his/her latitude of acceptance
 (rejection).
 A presented opinion is neither acceptable nor objectable if it is perceived to
 be neither close to or far from the individual's own stand point.
 This opinion is said to be in his/her latitude of noncommitment.
 Clearly, these three latitudes differ from person to person and they depend
 on factors such as individual's ego involvement and the person's familiarity
 of the subject of
 discussion~\cite{Eagly:1993,Sherif:1961,Sherif:1965,Griffin:2008}.
 When the presented opinion is in one's latitude of acceptance (rejection) or
 perhaps also in the nearby latitude of noncommitment, assimilation (contrast)
 occurs in the sense that the presented opinion is perceived to be closer to
 (farther from) one's stand point than it truly is.
 Moreover, this positively-evaluated (negatively-evaluated) opinion may cause
 the person to move his stand point towards (away from) it.
 The greater the difference between the individual's and the presented
 opinions, the more the resultant attitude change in general.
 The phenomenon of moving away from the presented opinion through contrast is
 called the boomerang
 effect~\cite{Eagly:1993,Griffin:2008,Sherif:1961,Sherif:1965}.
 The opinion change due to boomerang effect, however, is generally smaller
 than the opinion change induced by assimilation.
 Thus, not every psychological experiment unambiguously shows its
 existence~\cite{Griffin:2008}, making it perhaps the most controversial part
 of the \SJT.
 In fact, some social psychologists do not consider the boomerang effect to be
 one of the core thesis of \SJT and some even cast doubt on its
 existence~\cite{Eagly:1993}.
 Here we adopt the view that the boomerang effect is one of the central themes
 of \SJT whose effect, in general, is rather weak in comparison to the opinion
 change due assimilation.
 Finally, whenever the presented opinion is in the person's latitude of
 noncommitment which is not close to his/her latitudes of acceptance or
 rejection, then there is little chance for him/her to change his/her mind.
 Consequently, the most effective method to successfully persuade an individual
 is to present the opinion near the boundary of his/her latitudes of acceptance
 and noncommitment~\cite{Griffin:2008}.
 And just like most theories in social science, the above findings should be
 interpreted in statistical sense rather than as definitive rules governing
 every single persuasion and discussion~\cite{Eagly:1993, Griffin:2008,%
 Sherif:1961, Sherif:1965}.
 Thoroughly studied and advanced by social psychologists, \SJT is one of the
 most important theories in the field and is strongly supported by many
 psychological experiments especially concerning the latitudes of acceptance
 and noncommitment~\cite{Sherif:1961,Sherif:1965, Griffin:2008, Sakaki:1984,%
 Sarup:1991}.

 Recently, physical scientists entered this field by studying the more
 macroscopic aspects of the problem such as opinion formation and evolution in
 a social network using simple models and computer
 simulations~\cite{Sobkowicz:2009}.
 The variety of models proposed include the use of
 discrete or continuous opinions, discrete or continuous time, homogeneous or
 heterogeneous agents, fully connected or more realistic social
 networks~\cite{Sobkowicz:2009, Sznajd:2000, Deffuant:2000, Sousa:2005,%
 Deffuant:2002, Hegselmann:2002, Stauffer:2004a, Stauffer:2004b,%
 Jager:2005, Huet:2008, Lorenz:2010, Gandica:2010, Martins:2010,%
 MartinsKuba:2010, Kurmyshev:2011, Iniguez:2011, Xie:2011, Li:2011,%
 Sirbu:2013, Martins:2013, Crawford:2013}.
 Of particular importance is the continuous opinion agent-based model in a
 fully connected network introduced by Deffuant \emph{et al.} (\DW) with the
 feature that players only have latitudes of acceptance and
 noncommitment so that only the effect of assimilation is
 considered~\cite{Deffuant:2000, Deffuant:2002}.
 The appeal of this model is that it can be simulated efficiently by computers
 and its dynamics can be qualitatively understood.
 This model is also consistent with the social psychologists' finding that
 opinions can be reasonably well represented and measured as a
 continuum~\cite{Hegselmann:2002,Lodge:1981}.
 However, the absence of contrast and boomerang effect imply that \DW cannot be
 used to simulate opinion polarization in the real world in which opinions of
 the supporters of very different viewpoints become much more extreme.
 
 Various modifications of the \DW have been
 proposed~\cite{Deffuant:2002a, Amblard:2004, Jager:2005, Deffuant:2006, %
 Huet:2008, Lorenz:2010, Gandica:2010, Martins:2010, MartinsKuba:2010, %
 Kurmyshev:2011, Xie:2011, Li:2011, Sirbu:2013, Martins:2013, Crawford:2013}.
 To account for opinion polarization, some modified this model by introducing
 inflexible or contrarian
 players~\cite{Deffuant:2002a, Amblard:2004, Deffuant:2006, MartinsKuba:2010, %
 Xie:2011, Li:2011, Martins:2013},
 stochastic boomerang effect in the region of
 assimilation~\cite{Martins:2010, Sirbu:2013} and 
 vector-valued opinions~\cite{Sirbu:2013}.
 These models are not fully compatible with the \SJT as the agents' response in
 the latitude of rejection due to contrast are not properly treated.
 This is not ideal because in order to understand the macroscopic origin of
 opinion formation and polarization, one should combine the strengths of social
 psychology and physical science communities by introducing D-W-based models of
 opinion evolution whose rules are consistent with \SJT.
 In fact, this approach is beginning to gain acceptance among social
 psychologists~\cite{Mason:2007}.
 Actually, the only \SJT-based models we aware of are the ones proposed by
 Jager and Amblard (\JA)~\cite{Jager:2005} and its recent extension by Crawford
 \emph{et al.}~\cite{Crawford:2013} as well as the model of Huet
 \emph{et al.}~\cite{Huet:2008}.
 Jager and Amblard studied their model only by Monte Carlo simulation with
 very limited sample and agent sizes~\cite{Jager:2005}.
 The work of Crawford \emph{et al.} was more extensive, which included a simple
 analysis on eventual opinion distribution of the agents~\cite{Crawford:2013}.
 Note that both the models of Jager and Amblard~\cite{Jager:2005} and Crawford
 \emph{et al.}~\cite{Crawford:2013} involved agents with opinions on one issue
 only.
 In contrast, the model of Huet \emph{et al.}~\cite{Huet:2008} studied the
 response of agents based on their opinions on two issues by Monte Carlo
 simulation up to 5000~agents.
 
 While these works~\cite{Jager:2005, Crawford:2013, Huet:2008} point to the
 right direction, we argue in Sec.~\ref{Sec:Model} that the microscopic rules
 adopted in their models are questionable.
 Here we first proposed a minimalist \SJT-based model of opinion formation by
 extending the \JA in Ref.~\cite{Jager:2005}.
 This minimalist model is free of the questionable assumption implicitly used
 in Refs.~\cite{Jager:2005, Huet:2008, Crawford:2013}.
 Then in Secs.~\ref{Sec:Results} and~\ref{Sec:Ana}, we report that our
 minimalist model is simple enough to be studied both semi-analytically and
 numerically, and at the same time refined enough to show opinion polarization
 even in the case of homogeneous agents.
 By studying the agents' dynamics in Sec.~\ref{Sec:Ana}, we can understand the
 process of opinion clustering.
 In particular, using a simple mean-field analysis, we find that the most
 important parameters to determine the formation of extreme opinion clusters as
 well as the coexistence of both extreme and moderate opinion clusters are the
 values of two parameters $d_1$ and $d_2$ to be defined in Sec.~\ref{Sec:Model}
 which determine the widths of the regions for assimilation and boomerang
 effect to occur.
 Our analysis also shows that other factors such as network topology, agent's heterogeneity, and the
 detailed response dynamics due to assimilation and boomerang effect chiefly
 affect the opinion formation timescales.
 More importantly, we find in Sec.~\ref{Sec:Driving} that first-order phase
 transition in opinion clustering can occur occasionally when the widths of the
 assimilation and boomerang effect regions change very slowly.
 This shows that punctuated equilibrium in opinion distribution --- the
 observation that opinion distribution change often comes in a short burst
 between a long period of stasis, a notion first pointed out by Gould and
 Eldredge~\cite{Gould:1977} in evolution biology --- can occur even in a fully
 connected network, repudiating one of the criticisms~\cite{Tilcsik:2013} to
 the punctuated equilibrium theory in social science~\cite{Baumgartner:1993}.
 Finally, we give a brief outlook in Sec.~\ref{Sec:Outlook}.

\section{The model}
\label{Sec:Model}
 Just like the \DW~\cite{Deffuant:2000, Deffuant:2002} and the
 \JA~\cite{Jager:2005}, we consider a fixed connected network of $N$ agents
 each with a randomly and uniformly assigned initial opinion $x_i$ in a bounded
 interval, say, $[0,1]$.
 We call the opinions $0$ and $1$ extreme while those in between moderate.
 At each time step, we randomly pick two neighboring agents, say, $a$ and $b$,
 in the network and simulate their opinion changes after they meet and discuss
 by the following rules:
 \begin{itemize}
  \item \emph{The assimilation rule:} If $| x_a - x_b | < d_1$, then $x_a$ and
   $x_b$ are simultaneously updated as
   \begin{subequations} \label{E:d1_rule}
   \begin{align}
    x_a &\leftarrow x_a + \mu (x_b - x_a) , \\
    x_b &\leftarrow x_b + \mu (x_a - x_b) ,
   \end{align} 
   \end{subequations}
   where $\mu \in (0,0.5]$ is the convergence parameter.
  \item \emph{The boomerang effect rule:} If $| x_a - x_b | \ge d_2$, then
   $x_a$ and $x_b$ are simultaneously updated as
   \begin{subequations} \label{E:d2_rule}
   \begin{align}
    x_a &\leftarrow {\mathcal N}(x_a - \lambda (x_b - x_a)) , \\
    x_b &\leftarrow {\mathcal N}(x_b - \lambda (x_a - x_b)) ,
   \end{align} 
   \end{subequations}
   where $\lambda > 0$ is the divergence parameter, and
   \begin{equation}
    {\mathcal N}(x) =
     \begin{cases}
      x & \text{if } 0 \le x \le 1 , \\
      0 & \text{if } x < 0 , \\
      1 & \text{if } x > 1
     \end{cases}
    \label{E:normalization_def}
   \end{equation}
   is the normalization function which maps extreme opinions back to the range
   $[0,1]$.
  \item \emph{The neutral rule:}  The values of $x_a$ and $x_b$ do not change
   otherwise.
 \end{itemize}
 This serial opinion updating is repeated until the system is equilibrated.

 Clearly, our model is well-defined if $d_2 \ge d_1$ and is compatible with
 the \SJT with $d_1$ and $d_2$ reflecting the widths of the assimilation and
 boomerang effect regions, respectively.
 Furthermore, the rules are symmetric about $x = 1/2$.
 More importantly, our model is highly flexible.  Adapting it to model
 heterogeneous agents (in which each has different values of $d_1$ and $d_2$),
 different opinion change dynamics (by modifying
 Eqs.~\eqref{E:d1_rule}--\eqref{E:d2_rule} --- something that we are going to
 do in Sec.~\ref{Sec:Ana} below), and network topology are easy.

 Our model is very different from that of Huet \emph{et al.}~\cite{Huet:2008}
 since theirs is based on the repeated interaction of randomly picked pairs of
 agents whose responses are based on their opinion differences on two issues.
 Also, the most important difference between our model in the above form
 and the \JA model~\cite{Jager:2005} as well as its extension by Crawford
 \emph{et al.}~\cite{Crawford:2013} is that we use different convergence and
 divergence parameters $\mu$ and $\lambda$; while they set both to the same
 value.  Their choice is not very natural since $d_1 \le d_2$ would then imply
 the magnitude of opinion change due to boomerang effect must be greater than
 or equal to that due to assimilation, whose validity is not without
 doubt~\cite{Eagly:1993}.
 Note that both groups used Monte Carlo simulations to study their
 models~\cite{Jager:2005, Crawford:2013}.
 In fact, Jager and Amblard did not perform any analytical or semi-analytical
 study and Crawford \emph{et al.} only carried out a basic mean-field analysis
 which focused mainly on the asymptotic behavior rather than the detailed
 opinion dynamics of the agents.
 In contrast, our detailed mean-field analysis in Sec.~\ref{Sec:Ana} below
 shows that the dynamics of this type of models are so general that asymptotic
 behavior is very robust against any change in the assimilation and boomerang
 effect rules as well as the network topology provided that the average
 connectivity of the network is not too low.

% We have reservations on Huet \emph{et al.}'s model.
% In particular, we doubt if it is realistic for two randomly picked
% individuals, who may not know each other well, to discuss two controversial
% issues at a time.
% Besides, the individual's respond does not take into account the subtle
% psychological differences due to the order of discussion of the two issues.
 
\section{Simulation results}
\label{Sec:Results}
 We first present our findings for agents in a fully connected network with
 $\mu = 0.20$ and $\lambda =0.05$.
 These values are chosen to reflect the reality that agents generally have to
 interact several times before becoming extremists or sharing almost identical
 opinions.
 Besides, this choice makes sure that the magnitude of opinion change due to
 boomerang effect need not be greater than due to assimilation, which is
 consistent with findings of psychological experiments~\cite{Griffin:2008}.
 Since the network is fully connected, the dynamics of opinion distribution can
 be written as a master equation in the mean-field approximation.
 The master equation approach is computationally more efficient and generally
 more suitable to study the steady state opinion distribution than Monte Carlo
 method for a fully connected network~\cite{BenNaim:2003}.
 The results reported below are found by both Monte Carlo simulations and
 numerically solving the master equation.
 Both methods give similar results.

\begin{figure}%[t]
\begin{center}
\includegraphics[angle=0,width=0.9\columnwidth]{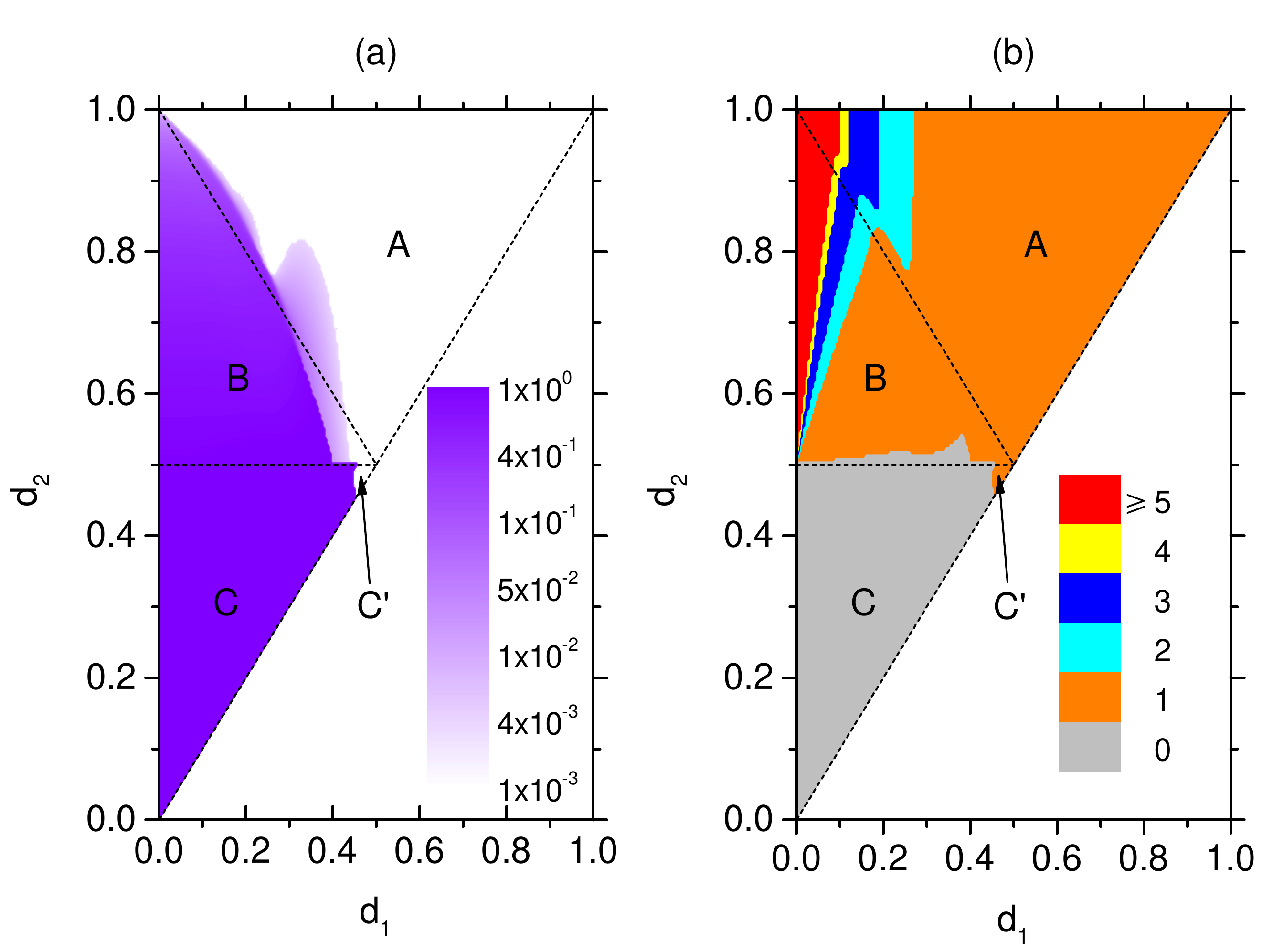}
\caption{
\label{fig:cluster_numbers}
 [Color online]
 (a)~Fraction of extreme opinion agents and (b)~number of moderate (major)
 opinion clusters in our model in a complete network found by numerically
 solving the master equation with the opinions divided into $1002$~bins for
 $\mu = 0.20$ and $\lambda = 0.05$.
}
\end{center}
\end{figure}

 By numerically solving the master equation, Fig.~\ref{fig:cluster_numbers}
 shows that the equilibrated opinion distribution for different values of $d_1$
 and $d_2$ can be divided into three regions.
 In region~A ($1 - d_2 \lesssim d_1$), the system evolves to clusters of
 moderate opinions similar to that of the \DW.
 In region~B ($d_1 \lesssim 1 - d_2 \lesssim 1/2$), the system equilibrates to
 two clusters of extreme opinions plus one or more clusters of moderate
 opinions similar to the \JA.
 And in region~C ($1 - d_2 \gtrsim 1/2$), the system evolves to two clusters of
 extreme opinions only.
 Again, this is similar to the results of the \JA.
 These findings are consistent with our Monte Carlo simulations except for the
 small region~C' in which $d_1,d_2 \lesssim 1/2$.
 We shall discuss this difference when we talk about the agents' dynamics
 below.

 Fig.~\ref{fig:cluster_numbers} also shows that the fraction of agents in an
 opinion cluster upon equilibration vary greatly for different values of $d_1$
 and $d_2$.
 In fact, Ben-Naim \emph{et al.} found that equilibrated opinion clusters of
 vastly different sizes can be present in the \DW.
 They called an opinion cluster with $\gg 10^{-3}$ ($\ll 10^{-3}$) fraction of
 agents a major (minor) cluster~\cite{BenNaim:2003}.
 Here, we clarify what an opinion cluster means in this paper.
 In our subsequent theoretical analysis, it refers to a connected subgraph of
 the network such that each agent in this subgraph has the same opinion.
 In addition, the ratio of agents in this subgraph to $N$ is non-zero in the
 large $N$ limit.
 Whereas in our Monte Carlo program, an opinion cluster is a connected subgraph
 of the network with at least $2\times 10^{-3}$ fraction of the agents such
 that opinion difference between any two agents in the subgraph is less than
 $d_1$ after equilibration.
 On the other hand, in our master equation program, consecutive discretized
 opinion bins each with fraction of opinion greater than a threshold of $2
 \times 10^{-3}$ upon equilibration is considered to be a cluster.
 In other words, unless otherwise stated, we do not consider minor opinion
 clusters in our simulations.

 Finally, we remark that we have tried several parameters pairs $(\mu,\lambda)$
 in our simulations and they all exhibit similar dynamics.
 In fact, our mean-field analysis in Sec.~\ref{Sec:Ana} shows why this is the
 case.

\section{Understanding our simulation results}
\label{Sec:Ana}
 Consider the following mean-field analysis.
 Let $y$ be the opinion of one of the agents chosen to interact at time $t =
 0$.
 Since the initial opinions are randomly and uniformly assigned, the net rate
 for $y$ to increase after the interaction equals
\begin{align}
 f(y) \equiv \ & \Prob(y \text{ increases}) - \Prob(y \text{ decreases})
  \nonumber \\
 = \ & \min (1 - y,d_1) + \max (y - d_2,0) - \min (y,d_1) \nonumber \\
 & \mathpunct{-} \max (1 - d_2 - y,0)
 \label{E:signed_movement_prob}
\end{align}
 for $0 < y < 1$.
 In addition, $f(0) = d_1 > 0$, $f(1) = -d_1 < 0$.
 There are three cases to consider.

\begin{figure}%[t]
\begin{center}
\includegraphics[angle=0,width=.9\columnwidth]{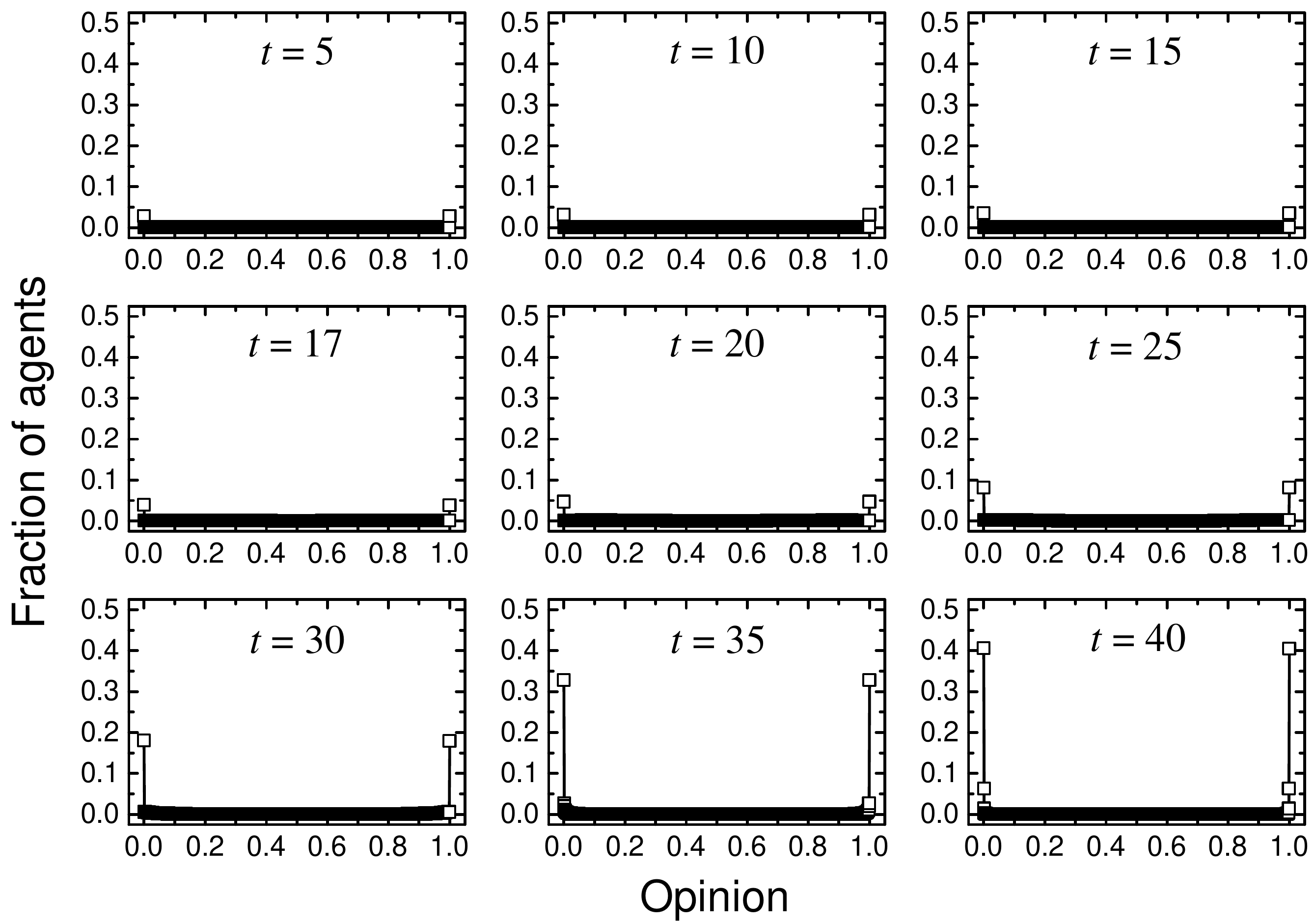}
\caption{
\label{fig:region_C_dynamics}
 Master equation solution of the opinion distribution at different time
 in region~C with $d_1=0.40$, $d_2=0.45$.  All other parameters are the same as
 those used in Fig.~\ref{fig:cluster_numbers}.
}
\end{center}
\end{figure}

 Case~(1): $d_2 < 1/2$, namely, most of the region~C.
 We only need to analyze the situation for opinions $x\in [1/2,1]$ as our model
 is symmetric about $x = 1/2$.
 Eq.~\eqref{E:signed_movement_prob} implies $f(x) > 0$ for $1/2 < x < 1$ and
 $f(1/2) = 0$.
 Thus, initially opinion tends to move towards $x = 1$; and $x = 1/2$ is an
 unstable equilibrium point.
 Whereas those with initial opinion $x = 1$ may change to an opinion in the
 range $R = [1- \mu d_1,1]$ after its first interaction due to the assimilation
 rule.
 Hence, opinions pile up around $R$ in the large $N$ limit shortly after $t =
 0$.
 Note that $f(y)$ is close to a linear function, increasing from $f(1/2) = 0$
 to $f(1-d_1) = 1-d_1-d_2$.
 So the number of agents with opinions around $1/2$ almost stays constant
 shortly after $t = 0$.
 Provided that $1-\mu d_1-1/2 \gtrsim d_1$, the assimilation rule has no effect
 between these piled up opinions in $R$ and those near $x = 1/2$.
 In this case, the net rate for opinion $x\in (1/2,1-\mu d_1)$ to increase at
 time $t \ge 1$ is greater than $f(x)$.
 More importantly, this positive feedback mechanism quickly kicks opinions out
 of $(1/2,1-\mu d_1)$.
 Finally, the assimilation rule among opinions in $R$ and the boomerang effect
 rule between opinions in $R$ and $[0,\mu d_1]$ assure that only the extreme
 opinions $x = 0$ and $1$ will be present in the long run.
 Fig.~\ref{fig:region_C_dynamics} as well as
 Videos~\ref{video:master_equation_dynamics}a
 and~\ref{video:Monte_Carlo_dynamics}a in the Supplemental Material~\cite{SM}
 show that this is indeed the observed dynamics in region~C.

 The situation is more complex when $d_1 \gtrsim 1/[2(1+\mu)]$ due to the competing
 dynamics of the assimilation and boomerang effect rules between opinions in
 $R$ and opinions near $x = 1/2$.
 Depending on the details of the dynamics, the master equation method finds
 that the assimilation rule may win resulting in a single moderate peak around
 $x = 1/2$; whereas the Monte Carlo method shows that this moderate peak may
 then be repelled to one of the extreme ends by a handful of remaining extreme
 opinion agents at the other end via the boomerang effect rule.
 (See Videos~\ref{video:master_equation_dynamics}b
 and~\ref{video:Monte_Carlo_dynamics}b as well as the discussions in
 the Supplemental Material~\cite{SM} on why the results of the two methods
 differ.)

 Note that one point is certain --- extreme and moderate opinion clusters
 cannot coexist for $d_2 < 1/2$.

\begin{figure}%[t]
\begin{center}
\includegraphics[angle=0,width=.9\columnwidth]{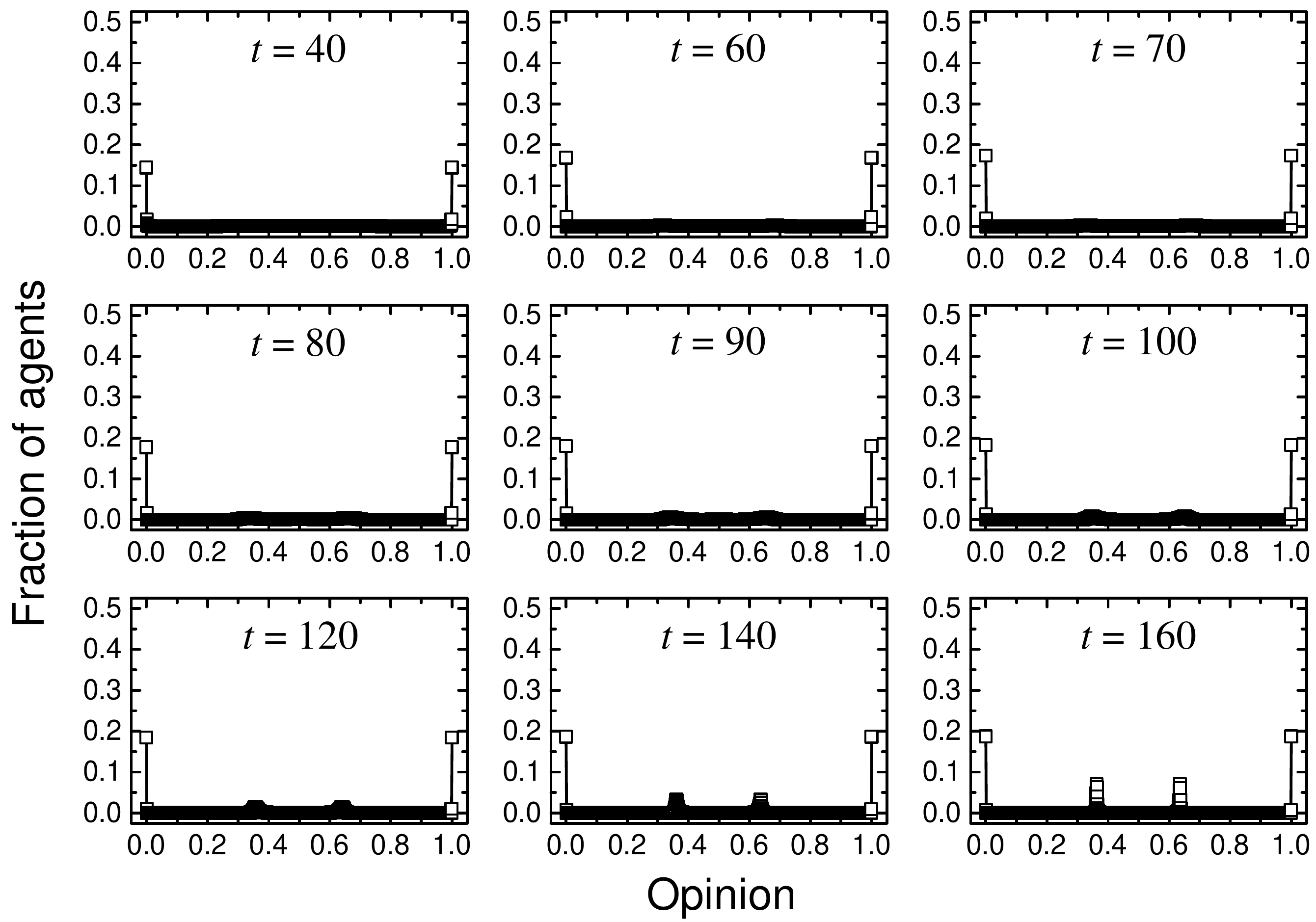}
\caption{
\label{fig:region_B_dynamics}
 Opinion dynamics in region~B with $d_1=0.12$, $d_2=0.80$.  All other
 parameters are the same as those used in Fig.~\ref{fig:region_C_dynamics}.
}
\end{center}
\end{figure}

 Case~(2): $d_1 < 1 - d_2 < 1/2$, that is, most of the region~B.
 Here Eq.~\eqref{E:signed_movement_prob} in the region becomes $f(x) > 0$ for
 $d_2 < x < 1$, and $f(x) = 0$ for $1/2 \le x \le d_2$.
 So we have a pile up of opinions in the interval $R$ and a migration of
 opinions from $R' = (d_2,1 - \mu d_1)$ to $R$ in the large $N$ limit shortly
 after $t = 0$.
 The same positive feedback mechanism acting on region~C then leads to the
 formation of the two extremist clusters provided that $d_1 \lesssim 1/[2(1+
 \mu)]$.
 Note that the opinion interval $(1-d_2,d_2)$ is in unstable equilibrium
 initially because local opinion clustering by the assimilation rule can grow.
 Besides, the depletion of opinions in $R'$ due to migration will in effect
 pull the opinions slightly less than $d_2$ to a lower value.
 These are precisely the effects governing the dynamics of the \DW.
 Thus, we end up with two extreme clusters plus several moderate ones as shown
 in Fig.~\ref{fig:cluster_numbers}.
 Moreover, the distance between two successive moderate (major) opinion
 clusters are separated by $\approx 2d_1$ in case of a fully connected
 network~\cite{Deffuant:2000, Deffuant:2002, BenNaim:2003} so that there are
 about $(2d_2-1)/2d_1$ of them.
 (See Fig.~\ref{fig:region_B_dynamics} as well as
 Videos~\ref{video:master_equation_dynamics}c--d
 and~\ref{video:Monte_Carlo_dynamics}c--d in the Supplemental
 Material~\cite{SM}.)

 There are two exceptions to this rule.
 Just like case~(1), if $d_1 \gtrsim 1/[2(1+\mu)]$, it is possible for opinions
 in $R$ to merge with opinions near $x = 1/2$ forming a single moderate cluster
 due to the assimilation rule.
 (See Fig.~\ref{fig:region_B_dynamics} as well as
 Videos~\ref{video:master_equation_dynamics}e
 and~\ref{video:Monte_Carlo_dynamics}e in the Supplemental Material~\cite{SM}.
 Unlike region~C', both master equation and Monte Carlo approaches give the
 same conclusion here.)
 Another situation is when $d_2 \approx 1/2$ so that the region $R'' = (1-d_2,
 d_2)$, where $f(x) = 0$, is very small.
 Depending on the details of the dynamics, the proportion of agents in $R''$
 may not be high enough to keep them in place before the region $R'$ is
 depleted.
 If this happens, the system will evolve to two extreme opinion peaks at $x =
 0$ and $1$; and this is what we find in Fig.~\ref{fig:cluster_numbers} as well
 as Videos~\ref{video:master_equation_dynamics}f
 and~\ref{video:Monte_Carlo_dynamics}f in the Supplemental Material~\cite{SM}.
 
\begin{figure}%[t]
\begin{center}
\includegraphics[angle=0,width=.9\columnwidth]{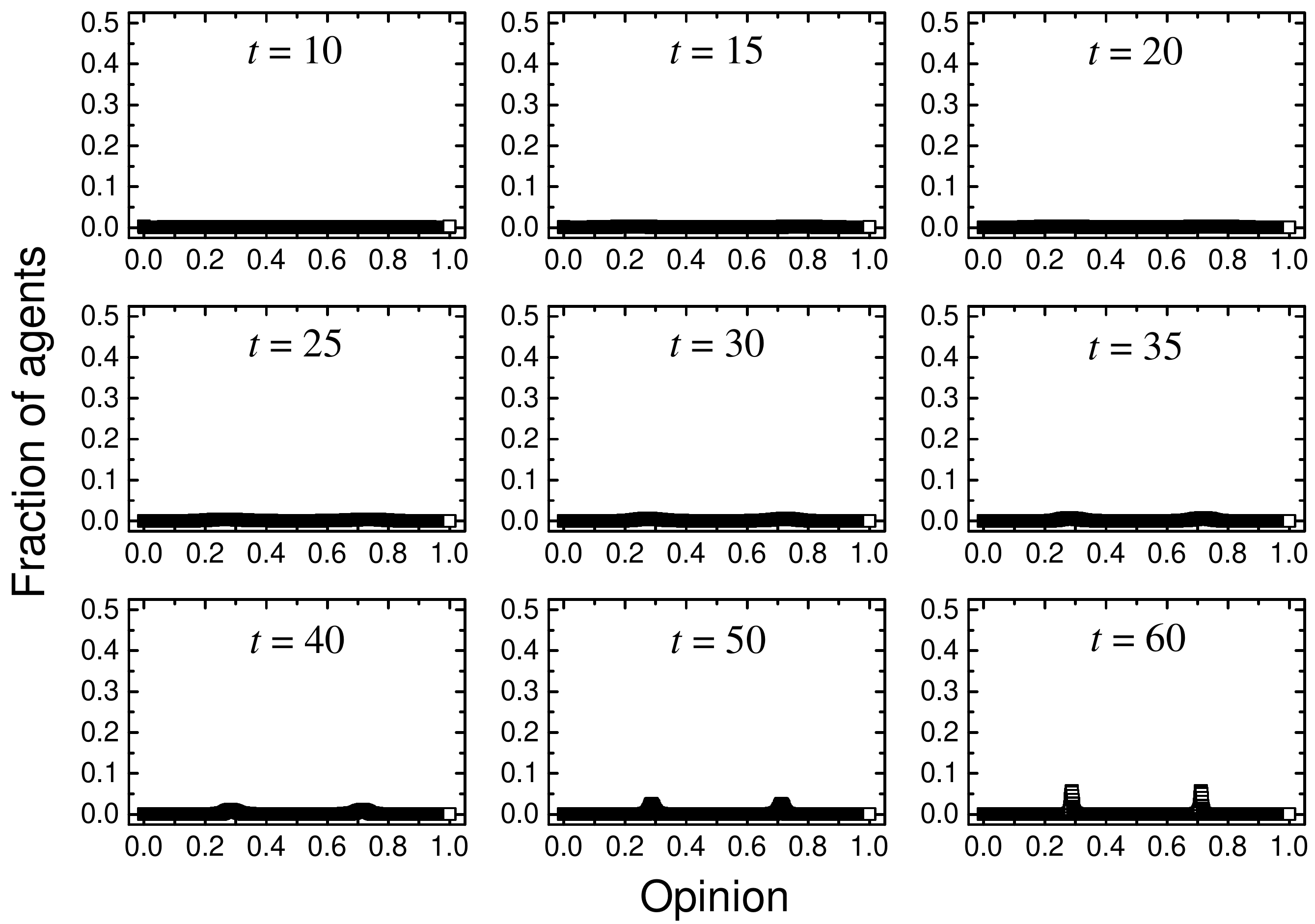}
\caption{
\label{fig:region_A_dynamics}
 Opinion dynamics in region~A with $d_1=0.25$, $d_2=0.90$.  All other
 parameters are the same as those used in Fig.~\ref{fig:region_C_dynamics}.
}
\end{center}
\end{figure}

 Case~(3): $1-d_2 < d_1$, namely, region~A and part of region~B.
 Here, $f(x) < 0$ for $\max (1 - d_1, 1/2) \le x < 1$, and $f(x) = 0$ for $1/2
 \le x \le \max( 1-d_1, 1/2)$.
 Hence, there is an initial migration of opinions from $(1 - d_1,1]$ to
 opinions around $x \lesssim 1-d_1$.
 Similar analysis in case~(2) shows that at least one moderate opinion cluster
 will form.
 (See Fig.~\ref{fig:region_A_dynamics} as well as
 Videos~\ref{video:master_equation_dynamics}g
 and~\ref{video:Monte_Carlo_dynamics}g in the Supplemental Material~\cite{SM}.)
 Nevertheless, there is a subtlety.
 If $1 - d_2$ is close to $d_1$ and $\lambda$ is sufficiently large, it is
 still possible for a small portion of agents to become extremists before they
 have time to join a moderate opinion cluster.
 The boundary between regions~A and~B, however, depends on the detailed
 dynamics of the system.
 Nevertheless, it is not possible to have extreme opinion peaks only in this
 case.
 (See Videos~\ref{video:master_equation_dynamics}h
 and~\ref{video:Monte_Carlo_dynamics}h in the Supplemental
 Material~\cite{SM}.)

\begin{figure}%[t]
\begin{center}
\includegraphics[angle=0,width=.45\columnwidth]{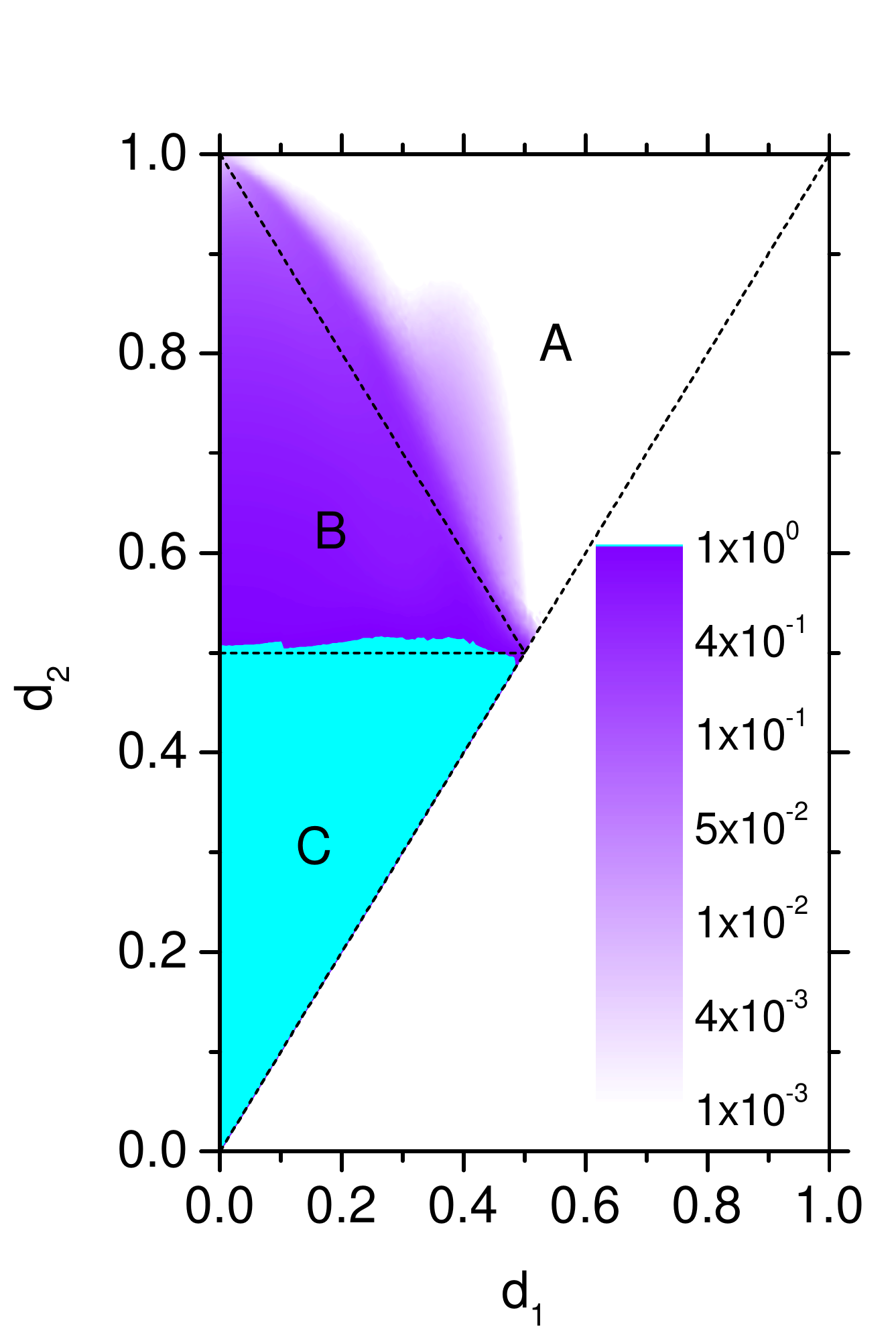}
\caption{[Color online]
 Phase diagram for our model in the \BA under different assimilation and
 boomerang effect rules described in the main text with $\mu' = 0.20$ and
 $\lambda' = 0.05$ obtained by averaging the results over 1000~independent
 Monte Carlo runs each with 1000~agents.  Region~C (extreme opinions only
 region) is colored cyan.  The portion of extreme opinion agents in region~B is
 represented by the intensity of purple color.
 \label{fig:extended_model_cluster_numbers}}
\end{center}
\end{figure}

 To summarize, we have argued that $d_1 + d_2 < 1$ and $d_2 > 1/2$ are the most
 important conditions for the formation of extreme and moderate opinion
 clusters, respectively.
 More importantly, the mean-field argument we have used does not depend on the
 precise form of the assimilation and boomerang effect rules.
 In fact, the key conditions used in our mean-field analysis are:
\begin{itemize}
 \item Agent's opinions can be described by a real number in $[0,1]$.
 \item All agents have the same $d_1$ and $d_2$.
 \item The assimilation (boomerang effect) rule makes the opinions of the two
  agents closer (farther) whereas the opinions are unchanged by the neutral
  rule.
 \item The criteria for applying the assimilation and boomerang effect rules
  are based only on the distance between two opinions $|x_a - x_b|$.
 \item The three rules governing the microscopic opinion change are symmetric
  about $x = 1/2$.  The last two conditions ensures that there is no prior bias
  toward one of the extreme opinions.
\end{itemize}
 In other words, the appeal of the above analysis is that conclusions can be
 drawn that are insensitive to factors such as network topology and the precise
 form of the agent dynamics as long as the average network connectivity is not
 too low and the agent dynamics is consistent with the \SJT.
 Indeed, Fig.~\ref{fig:extended_model_cluster_numbers} shows a similar phase
 diagram of our model in the Barab\'{a}si-Albert (B-A) scale-free
 network~\cite{Baraba:1999} even when
 the assimilation and boomerang effect rules for $x_a$ and $x_b$ in
 Eqs.~\eqref{E:d1_rule} and~\eqref{E:d2_rule} are changed to
\begin{subequations}\label{E:alt_d1_rule}
\begin{equation}
 x_a \leftarrow x_a + \frac{\mu' (x_b - x_a) (1 - |x_b - x_a|)}{d_1} ,
\end{equation}
\begin{equation}
 x_b \leftarrow x_b + \frac{\mu' (x_a - x_b) (1 - |x_b - x_a|)}{d_1}
\end{equation}
\end{subequations}
 and
\begin{subequations}\label{E:alt_d2_rule}
\begin{equation}
 x_a \leftarrow {\mathcal N} \left( x_a - \frac{\lambda' (x_b - x_a) (|x_b -
 x_a| - d_2)}{1 - d_2} \right) ,
\end{equation}
\begin{equation}
 x_b \leftarrow {\mathcal N} \left( x_b - \frac{\lambda' (x_a - x_b) (|x_b -
 x_a| - d_2)}{1 - d_2} \right) ,
\end{equation}
\end{subequations}
 respectively, where $\mu'$ and $\nu'$ are fixed positive parameters.
 Clearly, these modified assimilation and boomerang effect rules are very
 different from those used in the \DW and the \JA.
 More importantly, unlike our original rules in Eqs.~\eqref{E:d1_rule}
 and~\eqref{E:d2_rule}, the modified assimilation and boomerang effect rules
 are chosen so that agent's response is continuous across different latitudes,
 thereby demonstrating that the phase diagram is not sensitive to discontinuity
 in response across different latitudes.
 Note further that the number of moderate clusters in this case can be more
 than $\approx (2d_2-1)/2d_1$, which is consistent with the behavior of the
 \DW model in the \BA~\cite{Stauffer:2004a, Stauffer:2004b}.
 (See Videos~\ref{video:scale-free_dynamics}a--h in the Supplemental
 Material~\cite{SM}.)

 The major shortcoming in our mean-field analysis is that we cannot predict
 the height of each opinion cluster and the most likely location of each of
 them.
 Actually, by choosing $d_1$ and $d_2$ near the boundaries between regions~A,
 B and C, some of the equilibrated opinion clusters may contain less than 1\%
 of the population.

\section{Opinion dynamics of slowly driving $d_1$ and $d_2$}
\label{Sec:Driving}
 We go on to study the situation that the thresholds $d_1$ and $d_2$ in
 Eqs.~\eqref{E:d1_rule} and~\eqref{E:d2_rule} change to reflect the change in 
 the level of opinion tolerance in the society.
 While one's opinion may change by interacting with another agent once, it
 probably takes a much longer time for $d_1$ and $d_2$ to change since it
 reflects a fundamental change in the way the agents evaluate and response to
 the opinions of others.
 Here we consider the idealized situation that $d_1$ and $d_2$ change
 gradually in a timescale much longer than the opinion equilibration time of
 the system in a way analogous to the study of quasi-static equilibrium
 processes in thermodynamics.

 From the above analysis, we only need to consider the evolution of
 equilibrated opinions, which consists of extreme and/or moderate clusters,
 upon a small change in $d_1$ and $d_2$.
 Note that upon equilibration, the opinion difference between two adjacent
 agents belonging to two different opinion clusters must either be (i)~outside
 both the regions of assimilation and boomerang effect or (ii)~in the boomerang
 effect region and the two agents hold extreme opinions of $x = 0$ and $1$.
 Consequently, the opinion distribution will not change as one perturbs $d_1$
 and $d_2$ unless $d_1$ increases above or $d_2$ decreases below the opinion
 difference between two adjacent clusters.
 Some opinion clusters will merge in either cases.
 Thus, the opinion distribution stays constant most of the time and then
 suddenly change by opinion merging in a first-order phase transition.
 More importantly, a single moderate opinion cluster or two extreme opinion
 clusters are the only two stable fixed points of the system due to slowly
 random drifting of $d_1$ and $d_2$.

 Actually, opinion sudden changes in social issues after a long period of
 stasis, known as punctuated equilibria in social theory, is commonly
 observed~\cite{Baumgartner:1993}.
 Our analysis here shows that they may occur even in a fully connected social
 network, therefore repudiating the criticism by Tilcsik and
 Marquis~\cite{Tilcsik:2013}.
 In fact, punctuated equilibrium in our model originates from the separation of
 timescales between agents' interactions and the change in $d_1$ and $d_2$
 similar to the case of first-order phase transition in a quasi-statically
 evolving thermal system.

\section{Outlook}
\label{Sec:Outlook}
 In summary, we have proposed an agent-based and \SJT-compatible model by
 extending the works of Jager and Amblard~\cite{Jager:2005} and Crawford
 \emph{et al.}~\cite{Crawford:2013}.
 Our model can serve as a blueprint to study opinion formation dynamics.
 In our model, formation of extreme and/or moderate opinion clusters as well
 as punctuated equilibria are observed even in the case of homogeneous agents
 in a complete network.
 Besides, we identify the most important conditions for forming extreme and
 moderate opinion clusters by mean-field analysis.
 Further works should be done, including the addition of noise to the agent's
 response and a more detailed model of how $d_1$ and $d_2$ change, to make our
 model more realistic.
 Our next goal is to model opinion cluster splitting.

%\section*{Acknowledgments}

\bibliographystyle{apsrev4-1}

\bibliography{socphys1.5}

%merlin.mbs apsrev4-1.bst 2010-07-25 4.21a (PWD, AO, DPC) hacked
%Control: key (0)
%Control: author (72) initials jnrlst
%Control: editor formatted (1) identically to author
%Control: production of article title (-1) disabled
%Control: page (0) single
%Control: year (1) truncated
%Control: production of eprint (0) enabled
\begin{thebibliography}{38}%
\makeatletter
\providecommand \@ifxundefined [1]{%
 \@ifx{#1\undefined}
}%
\providecommand \@ifnum [1]{%
 \ifnum #1\expandafter \@firstoftwo
 \else \expandafter \@secondoftwo
 \fi
}%
\providecommand \@ifx [1]{%
 \ifx #1\expandafter \@firstoftwo
 \else \expandafter \@secondoftwo
 \fi
}%
\providecommand \natexlab [1]{#1}%
\providecommand \enquote  [1]{``#1''}%
\providecommand \bibnamefont  [1]{#1}%
\providecommand \bibfnamefont [1]{#1}%
\providecommand \citenamefont [1]{#1}%
\providecommand \href@noop [0]{\@secondoftwo}%
\providecommand \href [0]{\begingroup \@sanitize@url \@href}%
\providecommand \@href[1]{\@@startlink{#1}\@@href}%
\providecommand \@@href[1]{\endgroup#1\@@endlink}%
\providecommand \@sanitize@url [0]{\catcode `\\12\catcode `\$12\catcode
  `\&12\catcode `\#12\catcode `\^12\catcode `\_12\catcode `\%12\relax}%
\providecommand \@@startlink[1]{}%
\providecommand \@@endlink[0]{}%
\providecommand \url  [0]{\begingroup\@sanitize@url \@url }%
\providecommand \@url [1]{\endgroup\@href {#1}{\urlprefix }}%
\providecommand \urlprefix  [0]{URL }%
\providecommand \Eprint [0]{\href }%
\providecommand \doibase [0]{http://dx.doi.org/}%
\providecommand \selectlanguage [0]{\@gobble}%
\providecommand \bibinfo  [0]{\@secondoftwo}%
\providecommand \bibfield  [0]{\@secondoftwo}%
\providecommand \translation [1]{[#1]}%
\providecommand \BibitemOpen [0]{}%
\providecommand \bibitemStop [0]{}%
\providecommand \bibitemNoStop [0]{.\EOS\space}%
\providecommand \EOS [0]{\spacefactor3000\relax}%
\providecommand \BibitemShut  [1]{\csname bibitem#1\endcsname}%
\let\auto@bib@innerbib\@empty
%</preamble>
\bibitem [{\citenamefont {Eagly}\ and\ \citenamefont
  {Chaiken}(1993)}]{Eagly:1993}%
  \BibitemOpen
  \bibfield  {author} {\bibinfo {author} {\bibfnamefont {A.~H.}\ \bibnamefont
  {Eagly}}\ and\ \bibinfo {author} {\bibfnamefont {S.}~\bibnamefont
  {Chaiken}},\ }\href@noop {} {\emph {\bibinfo {title} {The Psychology Of
  Attitudes}}}\ (\bibinfo  {publisher} {Harcourt Brace College Publ.},\
  \bibinfo {address} {Philadelphia},\ \bibinfo {year} {1993})\BibitemShut
  {NoStop}%
\bibitem [{\citenamefont {{Also see, for example, Griffin,
  E.}}(2011)}]{Griffin:2008}%
  \BibitemOpen
  \bibfield  {author} {\bibinfo {author} {\bibnamefont {{Also see, for example,
  Griffin, E.}}},\ }\href@noop {} {\emph {\bibinfo {title} {A First Look At
  Communication Theory}}},\ \bibinfo {edition} {8th}\ ed.\ (\bibinfo
  {publisher} {McGraw-Hill},\ \bibinfo {address} {New York},\ \bibinfo {year}
  {2011})\ Chap.~\bibinfo {chapter} {15},\ \bibinfo {note} {and references
  cited therein}\BibitemShut {NoStop}%
\bibitem [{\citenamefont {Sherif}\ and\ \citenamefont
  {Hovland}(1961)}]{Sherif:1961}%
  \BibitemOpen
  \bibfield  {author} {\bibinfo {author} {\bibfnamefont {M.}~\bibnamefont
  {Sherif}}\ and\ \bibinfo {author} {\bibfnamefont {C.~I.}\ \bibnamefont
  {Hovland}},\ }\href@noop {} {\emph {\bibinfo {title} {Social Judgment:
  {A}ssimilation And Contrast Effects In Communication And Attitude Change}}}\
  (\bibinfo  {publisher} {Yale Univ. Press},\ \bibinfo {address} {New Haven,
  Connecticut},\ \bibinfo {year} {1961})\BibitemShut {NoStop}%
\bibitem [{\citenamefont {Sherif}\ \emph {et~al.}(1965)\citenamefont {Sherif},
  \citenamefont {Sherif},\ and\ \citenamefont {Nebergall}}]{Sherif:1965}%
  \BibitemOpen
  \bibfield  {author} {\bibinfo {author} {\bibfnamefont {C.~W.}\ \bibnamefont
  {Sherif}}, \bibinfo {author} {\bibfnamefont {M.}~\bibnamefont {Sherif}}, \
  and\ \bibinfo {author} {\bibfnamefont {R.~E.}\ \bibnamefont {Nebergall}},\
  }\href@noop {} {\emph {\bibinfo {title} {Attitude And Attitude Change: {T}he
  Social Judgment-Involvement Approach}}}\ (\bibinfo  {publisher} {Saunders},\
  \bibinfo {address} {Philadelphia},\ \bibinfo {year} {1965})\BibitemShut
  {NoStop}%
\bibitem [{\citenamefont {Sakaki}(1984)}]{Sakaki:1984}%
  \BibitemOpen
  \bibfield  {author} {\bibinfo {author} {\bibfnamefont {H.}~\bibnamefont
  {Sakaki}},\ }\href@noop {} {\bibfield  {journal} {\bibinfo  {journal}
  {Psychologia}\ }\textbf {\bibinfo {volume} {27}},\ \bibinfo {pages} {84}
  (\bibinfo {year} {1984})}\BibitemShut {NoStop}%
\bibitem [{\citenamefont {Sarup}\ \emph {et~al.}(1991)\citenamefont {Sarup},
  \citenamefont {Suchner},\ and\ \citenamefont {Gaylord}}]{Sarup:1991}%
  \BibitemOpen
  \bibfield  {author} {\bibinfo {author} {\bibfnamefont {G.}~\bibnamefont
  {Sarup}}, \bibinfo {author} {\bibfnamefont {R.~W.}\ \bibnamefont {Suchner}},
  \ and\ \bibinfo {author} {\bibfnamefont {G.}~\bibnamefont {Gaylord}},\ }\href
  {\doibase 10.2307/2786848} {\bibfield  {journal} {\bibinfo  {journal} {Social
  Psychol. Quarterly}\ }\textbf {\bibinfo {volume} {54}},\ \bibinfo {pages}
  {364} (\bibinfo {year} {1991})}\BibitemShut {NoStop}%
\bibitem [{\citenamefont {Sobkowicz}(2009)}]{Sobkowicz:2009}%
  \BibitemOpen
  \bibfield  {author} {\bibinfo {author} {\bibfnamefont {P.}~\bibnamefont
  {Sobkowicz}},\ }\href@noop {} {\bibfield  {journal} {\bibinfo  {journal} {J.
  Artificial Societies \& Social Simulation}\ }\textbf {\bibinfo {volume}
  {12}},\ \bibinfo {pages} {\#1, 11} (\bibinfo {year} {2009})},\ \bibinfo
  {note} {and references cited therein}\BibitemShut {NoStop}%
\bibitem [{\citenamefont {Sznajd-Weron}\ and\ \citenamefont
  {Sznajd}(2000)}]{Sznajd:2000}%
  \BibitemOpen
  \bibfield  {author} {\bibinfo {author} {\bibfnamefont {K.}~\bibnamefont
  {Sznajd-Weron}}\ and\ \bibinfo {author} {\bibfnamefont {J.}~\bibnamefont
  {Sznajd}},\ }\href {\doibase 10.1142/S0129183100000936} {\bibfield  {journal}
  {\bibinfo  {journal} {Int. J. Mod. Phys. C}\ }\textbf {\bibinfo {volume}
  {11}},\ \bibinfo {pages} {1157} (\bibinfo {year} {2000})}\BibitemShut
  {NoStop}%
\bibitem [{\citenamefont {Deffuant}\ \emph {et~al.}(2000)\citenamefont
  {Deffuant}, \citenamefont {Neau}, \citenamefont {Amblard},\ and\
  \citenamefont {Weisbuch}}]{Deffuant:2000}%
  \BibitemOpen
  \bibfield  {author} {\bibinfo {author} {\bibfnamefont {G.}~\bibnamefont
  {Deffuant}}, \bibinfo {author} {\bibfnamefont {D.}~\bibnamefont {Neau}},
  \bibinfo {author} {\bibfnamefont {F.}~\bibnamefont {Amblard}}, \ and\
  \bibinfo {author} {\bibfnamefont {G.}~\bibnamefont {Weisbuch}},\ }\href@noop
  {} {\bibfield  {journal} {\bibinfo  {journal} {Adv. Complex Sys.}\ }\textbf
  {\bibinfo {volume} {3}},\ \bibinfo {pages} {87} (\bibinfo {year}
  {2000})}\BibitemShut {NoStop}%
\bibitem [{\citenamefont {Sousa}(2005)}]{Sousa:2005}%
  \BibitemOpen
  \bibfield  {author} {\bibinfo {author} {\bibfnamefont {A.~O.}\ \bibnamefont
  {Sousa}},\ }\href {\doibase 10.1016/j.physa.2004.09.027} {\bibfield
  {journal} {\bibinfo  {journal} {Physica A}\ }\textbf {\bibinfo {volume}
  {348}},\ \bibinfo {pages} {701} (\bibinfo {year} {2005})}\BibitemShut
  {NoStop}%
\bibitem [{\citenamefont {Weisbuch}\ \emph {et~al.}(2002)\citenamefont
  {Weisbuch}, \citenamefont {Deffuant}, \citenamefont {Amblard},\ and\
  \citenamefont {Nadal}}]{Deffuant:2002}%
  \BibitemOpen
  \bibfield  {author} {\bibinfo {author} {\bibfnamefont {G.}~\bibnamefont
  {Weisbuch}}, \bibinfo {author} {\bibfnamefont {G.}~\bibnamefont {Deffuant}},
  \bibinfo {author} {\bibfnamefont {F.}~\bibnamefont {Amblard}}, \ and\
  \bibinfo {author} {\bibfnamefont {J.-P.}\ \bibnamefont {Nadal}},\ }\href
  {\doibase 10.1002/cplx.10031} {\bibfield  {journal} {\bibinfo  {journal}
  {Complexity}\ }\textbf {\bibinfo {volume} {7}},\ \bibinfo {pages} {55}
  (\bibinfo {year} {2002})}\BibitemShut {NoStop}%
\bibitem [{\citenamefont {Hegselmann}\ and\ \citenamefont
  {Krause}(2002)}]{Hegselmann:2002}%
  \BibitemOpen
  \bibfield  {author} {\bibinfo {author} {\bibfnamefont {R.}~\bibnamefont
  {Hegselmann}}\ and\ \bibinfo {author} {\bibfnamefont {U.}~\bibnamefont
  {Krause}},\ }\href@noop {} {\bibfield  {journal} {\bibinfo  {journal} {J.
  Artificial Societies \& Social Simulation}\ }\textbf {\bibinfo {volume}
  {5}},\ \bibinfo {pages} {\#3, 2} (\bibinfo {year} {2002})}\BibitemShut
  {NoStop}%
\bibitem [{\citenamefont {Stauffer}\ \emph {et~al.}(2004)\citenamefont
  {Stauffer}, \citenamefont {Sousa},\ and\ \citenamefont
  {Schulze}}]{Stauffer:2004a}%
  \BibitemOpen
  \bibfield  {author} {\bibinfo {author} {\bibfnamefont {D.}~\bibnamefont
  {Stauffer}}, \bibinfo {author} {\bibfnamefont {A.}~\bibnamefont {Sousa}}, \
  and\ \bibinfo {author} {\bibfnamefont {C.}~\bibnamefont {Schulze}},\
  }\href@noop {} {\bibfield  {journal} {\bibinfo  {journal} {J. Artificial
  Societies \& Social Simulation}\ }\textbf {\bibinfo {volume} {7(3)}},\
  \bibinfo {pages} {7} (\bibinfo {year} {2004})}\BibitemShut {NoStop}%
\bibitem [{\citenamefont {Stauffer}\ and\ \citenamefont
  {Meyer-Ortmanns}(2004)}]{Stauffer:2004b}%
  \BibitemOpen
  \bibfield  {author} {\bibinfo {author} {\bibfnamefont {D.}~\bibnamefont
  {Stauffer}}\ and\ \bibinfo {author} {\bibfnamefont {H.}~\bibnamefont
  {Meyer-Ortmanns}},\ }\href@noop {} {\bibfield  {journal} {\bibinfo  {journal}
  {Int. J. Mod. Phys. C}\ }\textbf {\bibinfo {volume} {15}},\ \bibinfo {pages}
  {241} (\bibinfo {year} {2004})}\BibitemShut {NoStop}%
\bibitem [{\citenamefont {Jager}\ and\ \citenamefont
  {Amblard}(2005)}]{Jager:2005}%
  \BibitemOpen
  \bibfield  {author} {\bibinfo {author} {\bibfnamefont {W.}~\bibnamefont
  {Jager}}\ and\ \bibinfo {author} {\bibfnamefont {F.}~\bibnamefont
  {Amblard}},\ }\href@noop {} {\bibfield  {journal} {\bibinfo  {journal} {Comp.
  Math. Organ. Theo.}\ }\textbf {\bibinfo {volume} {10}},\ \bibinfo {pages}
  {295} (\bibinfo {year} {2005})}\BibitemShut {NoStop}%
\bibitem [{\citenamefont {Huet}\ \emph {et~al.}(2008)\citenamefont {Huet},
  \citenamefont {Deffuant},\ and\ \citenamefont {Jager}}]{Huet:2008}%
  \BibitemOpen
  \bibfield  {author} {\bibinfo {author} {\bibfnamefont {S.}~\bibnamefont
  {Huet}}, \bibinfo {author} {\bibfnamefont {G.}~\bibnamefont {Deffuant}}, \
  and\ \bibinfo {author} {\bibfnamefont {W.}~\bibnamefont {Jager}},\
  }\href@noop {} {\bibfield  {journal} {\bibinfo  {journal} {Adv. Complex
  Sys.}\ }\textbf {\bibinfo {volume} {11}},\ \bibinfo {pages} {529} (\bibinfo
  {year} {2008})}\BibitemShut {NoStop}%
\bibitem [{\citenamefont {Lorenz}(2010)}]{Lorenz:2010}%
  \BibitemOpen
  \bibfield  {author} {\bibinfo {author} {\bibfnamefont {J.}~\bibnamefont
  {Lorenz}},\ }\href {\doibase 10.1002/cplx.20295} {\bibfield  {journal}
  {\bibinfo  {journal} {Complexity}\ }\textbf {\bibinfo {volume} {15}},\
  \bibinfo {pages} {43} (\bibinfo {year} {2010})}\BibitemShut {NoStop}%
\bibitem [{\citenamefont {Gandica}\ \emph {et~al.}(2010)\citenamefont
  {Gandica}, \citenamefont {del Castillo-Mussot}, \citenamefont {V{\'a}zquez},\
  and\ \citenamefont {Rojas}}]{Gandica:2010}%
  \BibitemOpen
  \bibfield  {author} {\bibinfo {author} {\bibfnamefont {Y.}~\bibnamefont
  {Gandica}}, \bibinfo {author} {\bibfnamefont {M.}~\bibnamefont {del
  Castillo-Mussot}}, \bibinfo {author} {\bibfnamefont {G.~J.}\ \bibnamefont
  {V{\'a}zquez}}, \ and\ \bibinfo {author} {\bibfnamefont {S.}~\bibnamefont
  {Rojas}},\ }\href {\doibase 10.1016/j.physa.2010.08.025} {\bibfield
  {journal} {\bibinfo  {journal} {Physica A}\ }\textbf {\bibinfo {volume}
  {389}},\ \bibinfo {pages} {5864} (\bibinfo {year} {2010})}\BibitemShut
  {NoStop}%
\bibitem [{\citenamefont {Martins}\ \emph {et~al.}(2010)\citenamefont
  {Martins}, \citenamefont {Pineda},\ and\ \citenamefont
  {Toral}}]{Martins:2010}%
  \BibitemOpen
  \bibfield  {author} {\bibinfo {author} {\bibfnamefont {T.~V.}\ \bibnamefont
  {Martins}}, \bibinfo {author} {\bibfnamefont {M.}~\bibnamefont {Pineda}}, \
  and\ \bibinfo {author} {\bibfnamefont {R.}~\bibnamefont {Toral}},\ }\href
  {\doibase 10.1209/0295-5075/91/48003} {\bibfield  {journal} {\bibinfo
  {journal} {Europhys. Lett.}\ }\textbf {\bibinfo {volume} {91}},\ \bibinfo
  {pages} {48003} (\bibinfo {year} {2010})}\BibitemShut {NoStop}%
\bibitem [{\citenamefont {Martins}\ and\ \citenamefont
  {Kuba}(2010)}]{MartinsKuba:2010}%
  \BibitemOpen
  \bibfield  {author} {\bibinfo {author} {\bibfnamefont {A.~C.~R.}\
  \bibnamefont {Martins}}\ and\ \bibinfo {author} {\bibfnamefont {C.~D.}\
  \bibnamefont {Kuba}},\ }\href@noop {} {\bibfield  {journal} {\bibinfo
  {journal} {Adv. Complex Sys.}\ }\textbf {\bibinfo {volume} {13}},\ \bibinfo
  {pages} {621} (\bibinfo {year} {2010})}\BibitemShut {NoStop}%
\bibitem [{\citenamefont {Kurmyshev}\ \emph {et~al.}(2011)\citenamefont
  {Kurmyshev}, \citenamefont {Juarez},\ and\ \citenamefont
  {Gonzalez-Silva}}]{Kurmyshev:2011}%
  \BibitemOpen
  \bibfield  {author} {\bibinfo {author} {\bibfnamefont {E.}~\bibnamefont
  {Kurmyshev}}, \bibinfo {author} {\bibfnamefont {H.~A.}\ \bibnamefont
  {Juarez}}, \ and\ \bibinfo {author} {\bibfnamefont {R.~A.}\ \bibnamefont
  {Gonzalez-Silva}},\ }\href@noop {} {\bibfield  {journal} {\bibinfo  {journal}
  {Physica A}\ }\textbf {\bibinfo {volume} {390}},\ \bibinfo {pages} {2945}
  (\bibinfo {year} {2011})}\BibitemShut {NoStop}%
\bibitem [{\citenamefont {I{\~n}iguez}\ \emph {et~al.}(2011)\citenamefont
  {I{\~n}iguez}, \citenamefont {Kert{\'e}sz}, \citenamefont {Kaski},\ and\
  \citenamefont {Barrio}}]{Iniguez:2011}%
  \BibitemOpen
  \bibfield  {author} {\bibinfo {author} {\bibfnamefont {G.}~\bibnamefont
  {I{\~n}iguez}}, \bibinfo {author} {\bibfnamefont {J.}~\bibnamefont
  {Kert{\'e}sz}}, \bibinfo {author} {\bibfnamefont {K.~K.}\ \bibnamefont
  {Kaski}}, \ and\ \bibinfo {author} {\bibfnamefont {R.~A.}\ \bibnamefont
  {Barrio}},\ }\href@noop {} {\bibfield  {journal} {\bibinfo  {journal} {Phys.
  Rev. E}\ }\textbf {\bibinfo {volume} {83}},\ \bibinfo {pages} {016111}
  (\bibinfo {year} {2011})}\BibitemShut {NoStop}%
\bibitem [{\citenamefont {Xie}\ \emph {et~al.}(2011)\citenamefont {Xie},
  \citenamefont {Sreenivasan}, \citenamefont {Korniss}, \citenamefont {Zhang},
  \citenamefont {Lim},\ and\ \citenamefont {Szymanski}}]{Xie:2011}%
  \BibitemOpen
  \bibfield  {author} {\bibinfo {author} {\bibfnamefont {J.}~\bibnamefont
  {Xie}}, \bibinfo {author} {\bibfnamefont {S.}~\bibnamefont {Sreenivasan}},
  \bibinfo {author} {\bibfnamefont {G.}~\bibnamefont {Korniss}}, \bibinfo
  {author} {\bibfnamefont {W.}~\bibnamefont {Zhang}}, \bibinfo {author}
  {\bibfnamefont {C.}~\bibnamefont {Lim}}, \ and\ \bibinfo {author}
  {\bibfnamefont {B.~K.}\ \bibnamefont {Szymanski}},\ }\href@noop {} {\bibfield
   {journal} {\bibinfo  {journal} {Phys. Rev. E}\ }\textbf {\bibinfo {volume}
  {84}},\ \bibinfo {pages} {011130} (\bibinfo {year} {2011})}\BibitemShut
  {NoStop}%
\bibitem [{\citenamefont {Li}\ \emph {et~al.}(2011)\citenamefont {Li},
  \citenamefont {Braunstein}, \citenamefont {Havlin},\ and\ \citenamefont
  {Stanley}}]{Li:2011}%
  \BibitemOpen
  \bibfield  {author} {\bibinfo {author} {\bibfnamefont {Q.}~\bibnamefont
  {Li}}, \bibinfo {author} {\bibfnamefont {L.~A.}\ \bibnamefont {Braunstein}},
  \bibinfo {author} {\bibfnamefont {S.}~\bibnamefont {Havlin}}, \ and\ \bibinfo
  {author} {\bibfnamefont {H.~E.}\ \bibnamefont {Stanley}},\ }\href@noop {}
  {\bibfield  {journal} {\bibinfo  {journal} {Phys. Rev. E}\ }\textbf {\bibinfo
  {volume} {84}},\ \bibinfo {pages} {066101} (\bibinfo {year}
  {2011})}\BibitemShut {NoStop}%
\bibitem [{\citenamefont {S{\^i}rbu}\ \emph {et~al.}(2013)\citenamefont
  {S{\^i}rbu}, \citenamefont {Loreto}, \citenamefont {Servedio},\ and\
  \citenamefont {Tria}}]{Sirbu:2013}%
  \BibitemOpen
  \bibfield  {author} {\bibinfo {author} {\bibfnamefont {A.}~\bibnamefont
  {S{\^i}rbu}}, \bibinfo {author} {\bibfnamefont {V.}~\bibnamefont {Loreto}},
  \bibinfo {author} {\bibfnamefont {V.~D.~P.}\ \bibnamefont {Servedio}}, \ and\
  \bibinfo {author} {\bibfnamefont {F.}~\bibnamefont {Tria}},\ }\href@noop {}
  {\bibfield  {journal} {\bibinfo  {journal} {J. Stat. Phys.}\ }\textbf
  {\bibinfo {volume} {151}},\ \bibinfo {pages} {218} (\bibinfo {year}
  {2013})}\BibitemShut {NoStop}%
\bibitem [{\citenamefont {Martins}\ and\ \citenamefont
  {Galam}(2013)}]{Martins:2013}%
  \BibitemOpen
  \bibfield  {author} {\bibinfo {author} {\bibfnamefont {A.~C.~R.}\
  \bibnamefont {Martins}}\ and\ \bibinfo {author} {\bibfnamefont
  {S.}~\bibnamefont {Galam}},\ }\href@noop {} {\bibfield  {journal} {\bibinfo
  {journal} {Phys. Rev. E}\ }\textbf {\bibinfo {volume} {87}},\ \bibinfo
  {pages} {042807} (\bibinfo {year} {2013})}\BibitemShut {NoStop}%
\bibitem [{\citenamefont {Crawford}\ \emph {et~al.}(2013)\citenamefont
  {Crawford}, \citenamefont {Brooks},\ and\ \citenamefont
  {Sen}}]{Crawford:2013}%
  \BibitemOpen
  \bibfield  {author} {\bibinfo {author} {\bibfnamefont {C.}~\bibnamefont
  {Crawford}}, \bibinfo {author} {\bibfnamefont {L.}~\bibnamefont {Brooks}}, \
  and\ \bibinfo {author} {\bibfnamefont {S.}~\bibnamefont {Sen}},\ }in\
  \href@noop {} {\emph {\bibinfo {booktitle} {Proceedings Of The 2013 Int.
  Conf. On Autonomous Agents And Multi-Agent Systems}}},\ \bibinfo {series and
  number} {AAMAS 2013},\ \bibinfo {editor} {edited by\ \bibinfo {editor}
  {\bibfnamefont {T.}~\bibnamefont {Ito}}, \bibinfo {editor} {\bibfnamefont
  {C.}~\bibnamefont {Jonker}}, \bibinfo {editor} {\bibfnamefont
  {M.}~\bibnamefont {Gini}}, \ and\ \bibinfo {editor} {\bibfnamefont
  {O.}~\bibnamefont {Shehory}}}\ (\bibinfo  {publisher} {Int. Found. for
  Autonomous Agents and Multiagent Systems},\ \bibinfo {address} {Richland, SC,
  USA},\ \bibinfo {year} {2013})\ pp.\ \bibinfo {pages}
  {1225--1226}\BibitemShut {NoStop}%
\bibitem [{\citenamefont {{See, for example, Lodge, M.}}(1981)}]{Lodge:1981}%
  \BibitemOpen
  \bibfield  {author} {\bibinfo {author} {\bibnamefont {{See, for example,
  Lodge, M.}}},\ }\href@noop {} {\emph {\bibinfo {title} {Magnitude Scaling:
  {Q}uantitative Measurement Of Opinions}}}\ (\bibinfo  {publisher} {{SAGE}
  Publ.},\ \bibinfo {address} {Beverly Hills, CA},\ \bibinfo {year}
  {1981})\BibitemShut {NoStop}%
\bibitem [{\citenamefont {Deffuant}\ \emph {et~al.}(2002)\citenamefont
  {Deffuant}, \citenamefont {Amblard}, \citenamefont {Weisbuch},\ and\
  \citenamefont {Faure}}]{Deffuant:2002a}%
  \BibitemOpen
  \bibfield  {author} {\bibinfo {author} {\bibfnamefont {G.}~\bibnamefont
  {Deffuant}}, \bibinfo {author} {\bibfnamefont {F.}~\bibnamefont {Amblard}},
  \bibinfo {author} {\bibfnamefont {G.}~\bibnamefont {Weisbuch}}, \ and\
  \bibinfo {author} {\bibfnamefont {T.}~\bibnamefont {Faure}},\ }\href@noop {}
  {\bibfield  {journal} {\bibinfo  {journal} {J. Artificial Societies \& Social
  Simulation}\ }\textbf {\bibinfo {volume} {5}},\ \bibinfo {pages} {\#4, 1}
  (\bibinfo {year} {2002})}\BibitemShut {NoStop}%
\bibitem [{\citenamefont {Amblard}\ and\ \citenamefont
  {Deffuant}(2004)}]{Amblard:2004}%
  \BibitemOpen
  \bibfield  {author} {\bibinfo {author} {\bibfnamefont {F.}~\bibnamefont
  {Amblard}}\ and\ \bibinfo {author} {\bibfnamefont {G.}~\bibnamefont
  {Deffuant}},\ }\href@noop {} {\bibfield  {journal} {\bibinfo  {journal}
  {Physica A}\ }\textbf {\bibinfo {volume} {343}},\ \bibinfo {pages} {725}
  (\bibinfo {year} {2004})}\BibitemShut {NoStop}%
\bibitem [{\citenamefont {Deffuant}(2006)}]{Deffuant:2006}%
  \BibitemOpen
  \bibfield  {author} {\bibinfo {author} {\bibfnamefont {G.}~\bibnamefont
  {Deffuant}},\ }\href@noop {} {\bibfield  {journal} {\bibinfo  {journal} {J.
  Artificial Societies \& Social Simulation}\ }\textbf {\bibinfo {volume}
  {9}},\ \bibinfo {pages} {\#3, 8} (\bibinfo {year} {2006})}\BibitemShut
  {NoStop}%
\bibitem [{\citenamefont {Mason}\ \emph {et~al.}(2007)\citenamefont {Mason},
  \citenamefont {Conrey},\ and\ \citenamefont {Smith}}]{Mason:2007}%
  \BibitemOpen
  \bibfield  {author} {\bibinfo {author} {\bibfnamefont {W.~A.}\ \bibnamefont
  {Mason}}, \bibinfo {author} {\bibfnamefont {F.~R.}\ \bibnamefont {Conrey}}, \
  and\ \bibinfo {author} {\bibfnamefont {E.~R.}\ \bibnamefont {Smith}},\
  }\href@noop {} {\bibfield  {journal} {\bibinfo  {journal} {Personality Soc.
  Psychol. Rev.}\ }\textbf {\bibinfo {volume} {11}},\ \bibinfo {pages} {279}
  (\bibinfo {year} {2007})}\BibitemShut {NoStop}%
\bibitem [{\citenamefont {Gould}\ and\ \citenamefont
  {Eldredge}(1977)}]{Gould:1977}%
  \BibitemOpen
  \bibfield  {author} {\bibinfo {author} {\bibfnamefont {S.~J.}\ \bibnamefont
  {Gould}}\ and\ \bibinfo {author} {\bibfnamefont {N.}~\bibnamefont
  {Eldredge}},\ }\href@noop {} {\bibfield  {journal} {\bibinfo  {journal}
  {Paleobiology}\ }\textbf {\bibinfo {volume} {3}},\ \bibinfo {pages} {115}
  (\bibinfo {year} {1977})}\BibitemShut {NoStop}%
\bibitem [{\citenamefont {Tilcsik}\ and\ \citenamefont
  {Marquis}(2013)}]{Tilcsik:2013}%
  \BibitemOpen
  \bibfield  {author} {\bibinfo {author} {\bibfnamefont {A.}~\bibnamefont
  {Tilcsik}}\ and\ \bibinfo {author} {\bibfnamefont {C.}~\bibnamefont
  {Marquis}},\ }\href@noop {} {\bibfield  {journal} {\bibinfo  {journal}
  {Admin. Sci. Quart.}\ }\textbf {\bibinfo {volume} {58}},\ \bibinfo {pages}
  {111} (\bibinfo {year} {2013})}\BibitemShut {NoStop}%
\bibitem [{\citenamefont {Baumgartner}\ and\ \citenamefont
  {Jones}(1993)}]{Baumgartner:1993}%
  \BibitemOpen
  \bibfield  {author} {\bibinfo {author} {\bibfnamefont {F.~R.}\ \bibnamefont
  {Baumgartner}}\ and\ \bibinfo {author} {\bibfnamefont {B.~D.}\ \bibnamefont
  {Jones}},\ }\href@noop {} {\emph {\bibinfo {title} {Agendas And Instability
  In American Politics}}}\ (\bibinfo  {publisher} {U of Chicago P},\ \bibinfo
  {address} {Chicago},\ \bibinfo {year} {1993})\BibitemShut {NoStop}%
\bibitem [{\citenamefont {Ben-Naim}\ \emph {et~al.}(2003)\citenamefont
  {Ben-Naim}, \citenamefont {Krapivsky},\ and\ \citenamefont
  {Redner}}]{BenNaim:2003}%
  \BibitemOpen
  \bibfield  {author} {\bibinfo {author} {\bibfnamefont {E.}~\bibnamefont
  {Ben-Naim}}, \bibinfo {author} {\bibfnamefont {P.~L.}\ \bibnamefont
  {Krapivsky}}, \ and\ \bibinfo {author} {\bibfnamefont {S.}~\bibnamefont
  {Redner}},\ }\href@noop {} {\bibfield  {journal} {\bibinfo  {journal}
  {Physica D}\ }\textbf {\bibinfo {volume} {183}},\ \bibinfo {pages} {190}
  (\bibinfo {year} {2003})}\BibitemShut {NoStop}%
\bibitem [{SM()}]{SM}%
  \BibitemOpen
  \href@noop {} {}\bibinfo {note} {{S}ee {S}upplemental {M}aterial for the
  videos showing the opinion dynamics in our model as well as the explanation
  of the difference between the master equation and {M}onte {C}arlo simulation
  results in region~C'.}\BibitemShut {Stop}%
\bibitem [{\citenamefont {Barab{\'a}si}\ and\ \citenamefont
  {Albert}(1999)}]{Baraba:1999}%
  \BibitemOpen
  \bibfield  {author} {\bibinfo {author} {\bibfnamefont {A.-L.}\ \bibnamefont
  {Barab{\'a}si}}\ and\ \bibinfo {author} {\bibfnamefont {R.}~\bibnamefont
  {Albert}},\ }\href@noop {} {\bibfield  {journal} {\bibinfo  {journal}
  {Science}\ }\textbf {\bibinfo {volume} {286}},\ \bibinfo {pages} {509}
  (\bibinfo {year} {1999})}\BibitemShut {NoStop}%
\end{thebibliography}%

\newpage

\appendix
\section*{Supplemental Material For ``Social Judgment Theory Based Model On
 Opinion Formation, Polarization And Evolution''}
\label{Sec:Supp}
 Videos~\ref{video:master_equation_dynamics}
 and~\ref{video:Monte_Carlo_dynamics} depict the dynamics of our model in a
 fully connected network in various regions of the parameter space $(d_1,d_2)$
 found by numerically solving the master equation and in typical runs of
 Monte Carlo simulation using $N = 1000$ agents, respectively.
 These two approaches give similar results except in region~C' (as shown in
 Videos~\ref{video:master_equation_dynamics}b
 and~\ref{video:Monte_Carlo_dynamics}b) where $d_1, d_2$ are slightly less than
 $1/2$.
 In this exceptional case, the master equation approach gives a single moderate
 peak at $x = 1/2$ in the steady state; while our Monte Carlo simulation shows
 that this moderate peak can be meta-stable.
 More precisely, after a long time, the moderate peak at $x=1/2$ sometimes move
 towards one of the extreme ends giving eventually a major extreme peak plus a
 very small minor extreme peak at the other end.
 In fact, by finite-size scaling analysis, our Monte Carlo simulation suggests
 that all the steady states in region~C' are made up of one major and one minor
 extreme peaks in the large $N$ limit.

 This discrepancy may be caused by the followings.
 For our numerical solution to the master equation, numerical truncation error
 and a long decay time of the meta-stable state may lead to a wrong conclusion.
 More importantly, as an approach that deals with the evolution of opinion
 distribution, the master equation approach fails to capture the dynamics of
 certain opinion distributions in our model.
 For example, consider the opinion distribution in which agents are of opinion
 $x=1/2$ almost surely and at the same time with a measure zero number of
 agents with opinion $x = 0$.
 (One may think of the system configuration in which there is only one agent
 with $x = 0$ and all the remaining $N-1$ agents has $x=1/2$.
 Then we take the limit $N\to +\infty$.)
 For $d_2 < 1/2$ and $d_1 \lesssim d_2$, this configuration will evolve to the
 steady state with one major opinion peak at $x = 1$ and the opinions of those
 agents with $x = 0$ are unchanged.
 It takes a long time for the system to evolve to this state though.
 Surely, this is what we will find by running the Monte Carlo simulation for a
 sufficiently long time.
 However, since the opinion distribution in this example is equal to that of
 the Dirac delta function at $x = 1/2$, the master equation approach will
 wrongly predicts that the system is already in steady state.
 However, we are not sure if this is the cause of the discrepancy because we do
 not know if a uniformly distributed initial opinion will almost surely evolve
 to a meta-stable state like the one in the above example.

 In addition, our Monte Carlo simulation is not without trouble in region~C'
 because the convergence time is too long to be computationally feasible to
 perform finite-size scaling when $d_1, d_2$ are very close to $1/2$.
 So we cannot rule out the existence of a single stable moderate peak at $x =
 1/2$ in the large $N$ limit when $d_1$ and $d_2$ are very close to $1/2$.

 Finally, Video~\ref{video:scale-free_dynamics} shows the dynamics of our model
 in the \BA with $N = 1000$ using different assimilation and boomerang effect
 rules.  It demonstrates that it has very similar dynamics as in the case of
 the complete network although the number of moderate peaks may differ.

\newpage
\begin{widetext}

\begin{video}
\caption{
 \label{video:master_equation_dynamics}
 {\color{red} Due to file size limit, all videos are not uploaded.  Please
 contact the first author at hfchau@hku.hk if you want a copy.}
 Videos showing the dynamics of our model in a fully connected network in
 different regions of the parameter space $(d_1,d_2)$ found by numerically
 solving the master equation with the opinions divided into $1002$~bins.
 In particular, Video~\ref{video:master_equation_dynamics}a shows the typical
 dynamics in region~C using parameters $(d_1,d_2) = (0.30,0.40)$;
 Video~\ref{video:master_equation_dynamics}b shows the dynamics in region~C'
 using parameters $(0.47,0.49)$;
 Video~\ref{video:master_equation_dynamics}c shows the dynamics in region~B
 using parameters $(0.10,0.85)$ which results in three moderate opinion
 clusters plus two extreme opinion clusters;
 Video~\ref{video:master_equation_dynamics}d shows the dynamics in region~B
 using parameters $(0.15,0.80)$ which results in two moderate opinion clusters
 plus two extreme opinion clusters;
 Video~\ref{video:master_equation_dynamics}e shows the dynamics in region~A
 with $d_1 + d_2 < 1$ using parameters $(0.42,0.55)$ which results in a single
 moderate opinion cluster;
 Video~\ref{video:master_equation_dynamics}f shows the dynamics in region~C
 with $d_2 > 1/2$ using parameters $(0.37,0.52)$;
 Video~\ref{video:master_equation_dynamics}g shows the dynamics in region~A
 using parameters $(0.25,0.90)$;
 Video~\ref{video:master_equation_dynamics}h shows the dynamics in region~B
 with $d_1 + d_2 > 1$ using parameters $(0.13,0.90)$.
}
\end{video}

\begin{video}
\caption{
 \label{video:Monte_Carlo_dynamics}
 The same as Videos~\ref{video:master_equation_dynamics}a--h except that
 these are typical runs of Monte Carlo simulations for a network of
 $1000$~agents.
}
\end{video}

\begin{video}
\caption{
 \label{video:scale-free_dynamics}
 The same as Videos~\ref{video:Monte_Carlo_dynamics}a--h but for the case of
 the \BA using different assimilation and contrast rules as mentioned in the
 main text.
}
\end{video}

\end{widetext}

\end{document}